\documentclass[sigconf,10pt,nonacm]{acmart}

\renewcommand\footnotetextcopyrightpermission[1]{} 
\usepackage{natbib,hyperref}
\usepackage[normalem]{ulem} 

\usepackage{microtype}
\usepackage{amsmath,amsfonts}
\usepackage{algorithmic}
\usepackage{graphicx}
\usepackage{textcomp}
\usepackage{xcolor}
\usepackage{multirow}

\usepackage{mathtools} 
\usepackage[switch]{lineno}

\usepackage{caption} 
                        
\usepackage{algorithmic}
\usepackage[linesnumbered,lined,boxed,ruled,vlined]{algorithm2e}
\newcommand{\lIfElse}[3]{\lIf{#1}{#2 \textbf{else}~#3}}

\definecolor{commentgreen}{RGB}{2,112,10}

\SetCommentSty{mycommfont}

\newlength{\commentWidth}
\usepackage{listings}

\definecolor{darkolivegreen}{rgb}{0.33, 0.42, 0.18}
\lstdefinestyle{cpp}{
language=C++,
basicstyle=\ttfamily\small,
keywordstyle=\color{blue},
commentstyle=\color{darkolivegreen},
stringstyle=\color{red},
showstringspaces=false,
breaklines=true,
frame=none,
numbers=left,
xleftmargin=0.4em,
numbersep=1pt,
numberstyle=\scriptsize\color{gray},
escapechar=|
}

\AtBeginDocument{%
  \providecommand\BibTeX{{%
    \normalfont B\kern-0.5em{\scshape i\kern-0.25em b}\kern-0.8em\TeX}}}

\begin{document}

\title{Improving Locality in Sparse and Dense Matrix Multiplications}

\author{Mohammad Mahdi Salehi Dezfuli}
\affiliation{
  \institution{McMaster University}            
  \city{Hamilton}
  \country{Canada}                    
}
\email{salehm32@mcmaster.ca}          

\author{Kazem Cheshmi}
\orcid{0000-0002-2968-5176}             
\affiliation{
  \institution{McMaster University}            
  \city{Hamilton}
  \country{Canada}                    
}
\email{cheshmi@mcmaster.ca}          

\begin{abstract}
Consecutive matrix multiplications are commonly used in graph neural networks and sparse linear solvers. These operations frequently access the same matrices for both reading and writing. While reusing these matrices improves data locality, it presents a challenge due to the irregular dependencies between iterations across the two multiplication operations. Existing fusion methods often introduce excessive synchronization overhead or overlapped computations with limited benefits. This paper proposes tile fusion, a runtime approach that fuses tiles of the two matrix-matrix multiplications, where at least one of the involved matrices is sparse. Tile fusion aims to improve data locality while providing sufficient workload for cores in shared-memory multi-core processors.
For a pair of matrix-matrix multiplications, tile fusion outperforms unfused baseline and MKL implementations with a geometric mean speedup of 1.97$\times$ 1.64$\times$, respectively, on multi-core CPUs. 
 
\end{abstract}

\maketitle


\section{Introduction}

Consecutive calls to matrix multiplications are the computational bottleneck in many scientific~\cite{o1980block} and machine learning~\cite{wang2019deep,fey2019fast} applications. Particularly this paper focuses on accelerating a pair of matrix multiplications, represented as an equation:
\begin{equation}
\label{eq:chain}
   D = A  (B  C) 
\end{equation}
 where matrix $\underset{n \times n}A$ is  sparse, $\underset{n \times bCol}B$ is either sparse or dense, and $\underset{bCol \times cCol}C$ is dense. For example, in a layer of graph convolution network~\cite{kipf2016semi}, either cases happen. 
Existing frameworks such as PyTorch Geometric (PyG)~\cite{fey2019fast} and Deep Graph Library (DGL)~\cite{wang2019deep} break the expression into two matrix multiplication operations, $D_1 = B C$ and $D = A D_1$. The two operations are commonly mapped to a pair of General Matrix Multiplication (GeMM)-Sparse Matrix-Matrix Multiplication (SpMM) or SpMM-SpMM routines when $B$ is dense and sparse, respectively. These routines benefit from efficient tiling and load balancing techniques~\cite{hong2019adaptive,wang2014intel} that enable using memory and computing resources efficiently. However, $D_1$ is shared between the two routines and often a large matrix that can be reused but it is not used when the operation is mapped to GeMM or SpMM separately. 

Fusing operations or loops are commonly used to remove intermediate matrices between the two operations. Tensor compilers~\cite{kjolstad2017tensor,dias2022sparselnr,COMET:LCPC-20} generate a fused code for Equation~\ref{eq:chain} when $A$ is sparse and $B$ and $C$ are dense. The generated code iterates over $A$ and performs a general matrix-vector multiplication (GeMV) for each nonzero of $A$. While this removes the need for storing intermediate results, i.e. $D_1$, it causes random access to $B$ and thus inefficient use of memory hierarchy. Additionally, this methodology does not apply when $A$ and $B$ are sparse because memory accesses are unknown at compile time.

\begin{figure}[!t]
    \centering
    \includegraphics[width=0.9\linewidth]{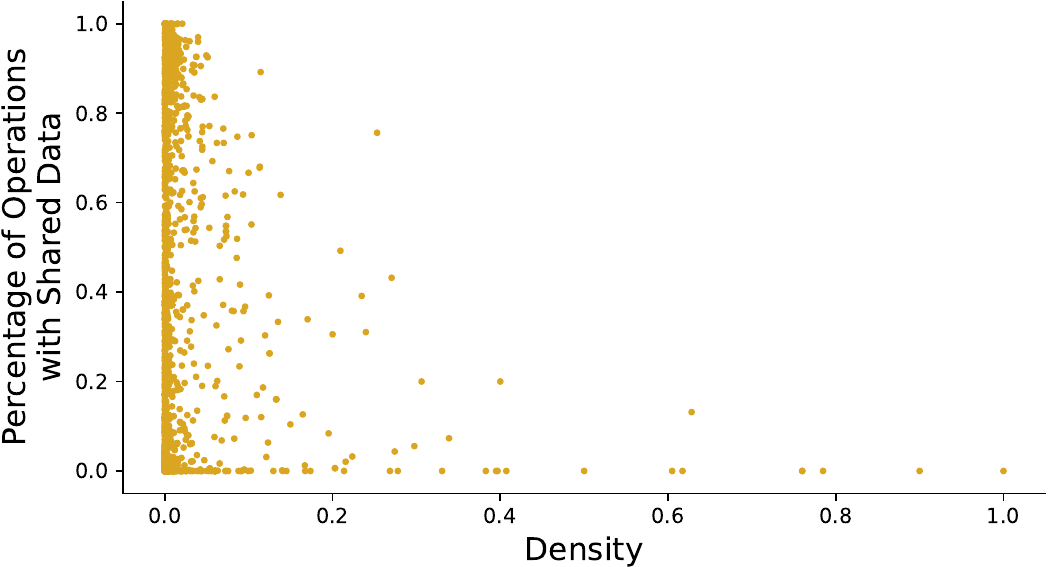}
    \caption{The ratio of computations in coarse fused tiles for all matrices from SuiteSparse for GEMM-SpMM operation.}
    \label{fig:fused-ratio}
\end{figure}

\begin{figure*}[!ht]
    \centering
    \includegraphics[width=\linewidth]{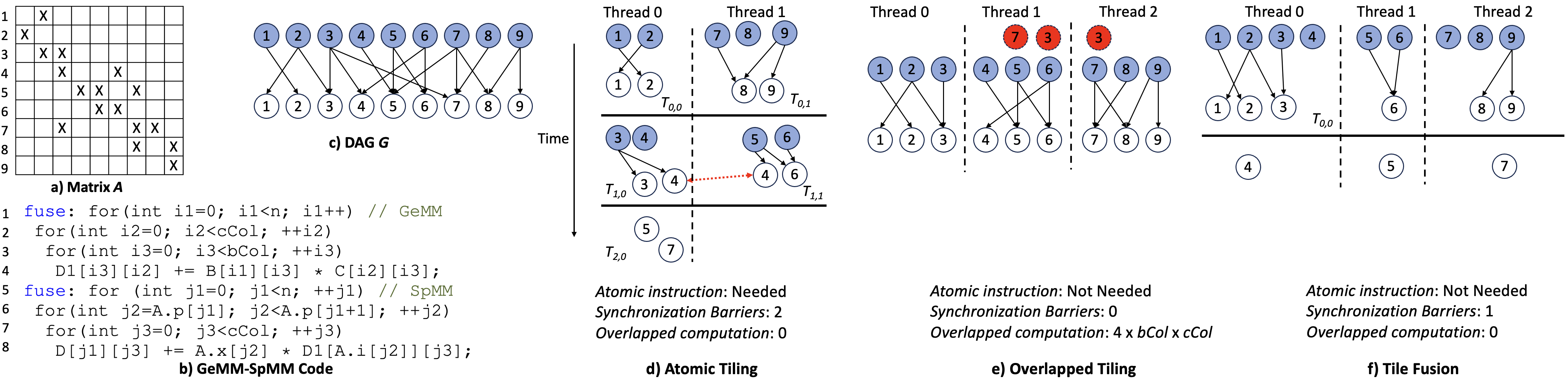}
    \caption{Three different iteration fusion schedules (Figure~\ref{fig:motivation}d--f) for the GeMM-SpMM in Figure~\ref{fig:motivation}b and the matrix in Figure~\ref{fig:motivation}a. Figure~\ref{fig:motivation}c shows the dependence DAG between iterations of the outermost loop of GeMM and SpMM, where colored and white vertices correspond to GeMM and SpMM iterations, respectively. Dark solid lines show synchronization barriers, the dotted red line shows a potential race condition, and vertical dashed lines show per thread workload.  }
    \label{fig:motivation}
\end{figure*}

Prior approaches such as sparse tiling~\cite{krieger2013loop} and communication-avoiding (CA)~\cite{demmel2008avoiding} methods have used sparsity information at runtime to fuse sparse matrix-vector multiplications (SpMV) and enable reuse between the two operations. 
They model SpMV operations as an iteration data acyclic graph (DAG) where vertices are iterations of the outermost loop of SpMV and edges represent dependencies between iterations. Then a scheduler tiles iterations of operations by grouping vertices of DAG at runtime. 
Then, sparse tiling uses barrier and atomic operations to ensure dependence between tiles are not violated during parallel execution. Some CA methods~\cite{demmel2008avoiding} replicate dependent iterations within a tile to make all tiles independent so they run in parallel without synchronization. 
Since GeMM, SpMM, and SpMV have parallel iterations in their outermost loops, the same techniques can be adopted for fusing GeMM-SpMM and SpMM-SpMM. However, the computation of each fused iteration in the two operations, is proportional with $bCol$ and $cCol$, increasing race conditions in sparse tiling and redundant computation in CA methods.

Coarse-grain tiles provide opportunities for fusion in sparse matrices and graphs without redundant computation or excessive synchronization. A coarse tile contains large enough iterations of the first operation such that it allows running some iterations of the second operation that solely depend on iterations inside the tile. This allows tiles to execute in parallel without synchronization.  Figure~\ref{fig:fused-ratio} shows the percentage of GeMM-SpMM computations that share data across the operations if coarse-grain tiles with the size of 2048 are selected for all 2893 matrices from SuiteSparse matrix collection~\cite{davis2011university}. As shown, an average of 34\% of GeMM-SpMM computation reuse data in fused coarse tiles.
However, growing the tiles reduces the number of parallel workloads, affecting load balance. Also, picking coarse grain tiles groups a larger number of iterations from the two operations. This grouping improves locality if the memory accesses of the tile fit within the size of the fast memory.

We propose sparsity-oriented tile fusion, in short, \textit{tile fusion}, that creates fused tiles based on the opportunities shown in Figure~\ref{fig:fused-ratio} to improve locality in GeMM-SpMM and SpMM-SpMM for shared memory multicore processors. 
This paper makes the following contributions:
\begin{itemize}
    \item Tile fusion scheduler and fused code that turn data reuse between and across iterations of GeMM and SpMM into locality. The tile fusion scheduler uses the sparsity pattern of $A$ and selects tile sizes and a number of tiles to ensure locality and load balance. 
    \item An implementation that is tested for a wide range of graphs and matrices and provides a speedup of 1.97$\times$ and 3.52$\times$ compared to existing unfused and best-fused codes. Also an analysis and adoption of prior tiling approaches and comparison with tile fusion. 
\end{itemize}

\section{Motivating Example}
We use the matrix in Figure~\ref{fig:motivation}a to discuss how different fusion strategies improve locality for computing Equation~\ref{eq:chain}. The corresponding code to the computation is shown in Figure~\ref{fig:motivation}b where lines~1--4 perform GeMM, $D_1 = B  C$, and lines~5--8 perform SpMM, $D = A D_1$. Iterations of loops \texttt{i1}  and \texttt{j1} are independent so they execute in parallel. Fusing loops \texttt{i1} and \texttt{j1} can potentially enable reusing $D_1$ but each iteration in \texttt{j1} depends on a variant number of \texttt{i1} iterations. This irregular dependence is due to \texttt{D1[A.i[j2]][j3]} in line~8 in Figure~\ref{fig:motivation}b, stemming from sparsity pattern of $A$. The DAG shown in Figure~\ref{fig:motivation}c shows the dependence between \texttt{i1} and \texttt{j1}. Colored and white vertices in Figure~\ref{fig:motivation}c represent iterations of \texttt{i1} and \texttt{j1} loops, respectively. Edges show dependence between iterations. While grouping vertices with common edges as a tile improves locality, dependence between tiles can prevent keeping all cores busy. 
Three different fused schedules of iterations for the DAG shown in Figure~\ref{fig:motivation}c are shown in Figure~\ref{fig:motivation}d--f for a processor with three cores.

Figure~\ref{fig:motivation}d shows five tiles composed of vertices of both computations with common edges.  Dependent tiles are separated by synchronization barriers to ensure partial order. Tiles are atomic to prevent race conditions.  
 For example, tile $\mathcal{T}_{1,0}$ and $\mathcal{T}_{1,1}$ depend on tile $\mathcal{T}_{0,0}$ and $\mathcal{T}_{0,1}$ thus a synchronization is needed between them.
 Iteration \texttt{j=4} is split among tiles $\mathcal{T}_{1,0}$ and $\mathcal{T}_{1,1}$, writing to the same location of \texttt{C}, thus an atomic operation is needed. The race condition is shown with the dotted red line in Figure~\ref{fig:motivation}. 
%
This schedule is inspired by sparse tiling~\cite{krieger2013loop} and named atomic tiling due to atomic operations used in tiles.
The chance of race condition on writing to \texttt{C} increases as the number of columns in $B$ nad $C$ increases.

Figure~\ref{fig:motivation}e shows overlapped tiles that create independent tiles by replicating dependent iterations. Replicated iterations are shown with red vertices in two tiles in Figure~\ref{fig:motivation}e. Therefore all fused tiles execute in parallel with no synchronization. 
Each replicated vertex in the tile corresponds to an iteration \texttt{i1} which multiplies a row of $B$ with $C$.  Therefore redundant computations increase with the number of columns in $B$ and $C$. Due to replicated iterations, this method is called overlapped tiling, inspired by CA~\cite{demmel2008avoiding} methods.

The tile fusion schedule is shown in Figure~\ref{fig:motivation}f where two groups of tiles are created, fused tiles and tiles of the SpMM iterations separated by one synchronization barrier. As shown, tiles in the schedule can be large, such as tile $\mathcal{T}_{0,0}$, to enable fusing more SpMM iterations, benefiting from coarse tile fusion shown in Figure~\ref{fig:fused-ratio}. The tiles contain a variable number of iterations to ensure the memory accesses of the tile remain local to the fast memory. Also, both levels have three independent workloads for all three cores. 
As a result of tile fusion, the performance of GeMM-SpMM for a subset of SuiteSparse matrices on a 20-core processor is faster than atomic tiling, overlapped tiling, and unfused code with a geometric mean of 13.6$\times$, 3.5$\times$, and 1.64$\times$, respectively.

\section{Tile Fusion}
Tile fusion looks into the sparsity pattern of the input matrix $A$ in GeMM-SpMM or SpMM-SpMM and creates a fused schedule. The tile fusion approach has an iteration scheduler and fused code.  
Both pairs of operations are commonly used in applications where the sparsity pattern of matrix $A$ and $B$ (when sparse) remains static during the execution. Therefore the created schedule will be computed once based on their sparsity and reused for the rest of the computation. The rest of this section explains how the scheduler computes the fused schedule and how codes are fused.

\begin{algorithm} [tb]

\SetAlgoLined
\DontPrintSemicolon
\SetAlgoLined
\begin{small}
\caption{\label{alg:scheduler}Tile Fusion Scheduler }
\SetKwInOut{Input}{Input}
\SetKwInOut{Output}{Output}
\Input{\,$G$, $bCol$, $cCol$, $p$, $cacheSize$, $ctSize$}
\Output{\,$\mathcal{T}$}
\tcc{Step 1: Coarse Tile Fusion}
$I \leftarrow range(rows(G))$\;
$J \leftarrow range(cols(G))$\;
\lIfElse{$\lceil|I|/ctSize\rceil \geq p$}{$t \leftarrow ctSize$}{$t \leftarrow \lceil|I|/p\rceil$}\label{lin:alg:sched:step1begin}
$\mathcal{F} \leftarrow (\{\},\{\})$\;
\For{ $i \in I$}{\label{lin:alg:sched:step1iloop}
    $v \leftarrow i/t$\;
    $\mathcal{F}_{0,v} \leftarrow \mathcal{F}_{0,v} \cup range(i, i+t)$\; \label{lin:alg:sched:step1range}
    \For{$j \leftarrow i \quad to \quad i+t \quad \land \quad j \in J $}{ \label{lin:alg:sched:step1jloop}
        \lIf{($i < inEdges(G,j) < i+t$)}{ \label{lin:alg:sched:step1correctness}
            $\mathcal{F}_{0,v} \leftarrow \mathcal{F}_{0,v} \cup j$ \label{lin:alg:sched:step1wf0}
        } \lElse{
            $\mathcal{F}_{1,v} \leftarrow \mathcal{F}_{1,v} \cup j$ \label{lin:alg:sched:step1wf1}
        }
        $j \leftarrow j+1$\;
    }
    $i \leftarrow i + t$\;
} 
$\mathcal{F}_{1,v} \leftarrow balance(\mathcal{F}_{1,v}, t)$\;\label{lin:alg:sched:step1end}
\tcc{Step 2: Fused Tile Splitting}
\For{$w \leftarrow 0 \; to \; 2$}{\label{lin:alg:sched:step2beg}
    \For{$v \leftarrow 0 \; to \;  |\mathcal{F}_{w}| $}{
            \lIf{$cost(\mathcal{F}_{w,v}, bCol, cCol)  > cacheSize$ }{\label{lin:alg:sched:step2cost}
                $\mathcal{T}_w \leftarrow \mathcal{T}_w \cup split(\mathcal{F}_{w,v}, bCol, cCol, cacheSize) $\label{lin:alg:sched:step2split}
            }\lElse{
                $\mathcal{T}_w \leftarrow \mathcal{T}_w \cup  \mathcal{F}_{w,v}$) \label{lin:alg:sched:step2nosplit}
            }
            $v \leftarrow v+1$\;
    }
    $w \leftarrow w+1$\;
}\label{lin:alg:sched:step2end}
\end{small}
\end{algorithm}

\subsection{Scheduler}
The tile fusion scheduler is shown in Algorithm~\ref{alg:scheduler} where it creates a schedule of fused tiles based on sparsity pattern at runtime. This subsection explains inputs, output, objective, and the two steps of the algorithm.

\subsubsection*{\textbf{Inputs and output}}
The tile fusion scheduler takes the DAG $G$, number of columns of $B$ and $C$ as $bCol$ and $cCol$, architecture-specific information $p$ and $cacheSize$, and heuristic parameter $ctSize$ as inputs and creates a set of fused tiles $\mathcal{T}$.
DAG $G$ represents dependence between $n$ iterations of the two fused loops where $G_{i,j}=1$ shows iteration $j$ of the second loop depends on the iteration $i$ of the first loop. 
$p$ represents the number of physical cores and $cacheSize$ shows the total size of caches per core in the target architecture. Each tile in the output fused tile $\mathcal{T}$ is shown with  $\mathcal{T}_{w,v}$ where $w$ and $v$ are \textit{wavefront} number and tile number, respectively. Wavefront is a set of iterations that can execute in any order without violating correctness. 

\subsubsection*{\textbf{Objective and Constraints}}
The objective of the tile fusion scheduler is to maximize the fused ratio across all tiles $\mathcal{T}$ while tiles fit into fast memory and only two wavefronts are allowed. The fused ratio is computed as total iterations of the second computation in the first wavefront over total iterations as shown in Equation~\ref{eq:fusedratio}:
\begin{equation}
\label{eq:fusedratio}
    fused\_ratio = \frac{\sum_{v=0}^{|\mathcal{T}_0|} |J_{0v}|}{|I| + |J|}
\end{equation}
where $J_{wv}$ represents the list of iterations from the second operation in tile $\mathcal{T}_{wv}$ and $I$ and $J$ shows the list of all iterations (or iteration space) of the first and second operations, respectively. $\mathcal{T}_0$ is the set of tiles in the first wavefront. Operator $|.|$ shows the cardinal number for a set or size of a list. 
The tile fusion scheduler maximizes fused ratio under two constraints, \textit{load balance constraint} and \textit{locality constraint}. The load balance constraint ensures the schedule has a maximum of two synchronization barriers (or two wavefronts) and the number of tiles in each wavefront is larger than the number of cores, i.e., $\forall 0 \leq w < 2; \quad |\mathcal{T}_{w}| \geq p$. The locality constraint ensures the data movement cost for a tile is smaller than cache sizes for each core ($cacheSize$). In other words the $\forall w,v;\quad cost(\mathcal{T}_{w,v}) < cacheSize$ where $cost(\mathcal{T}_{w,v})$ is data movement cost of $T_{wv}$. 


Figure~\ref{fig:motivation}c shows an example DAG $G$. There is an edge from the first iteration of GeMM to the second iteration of SpMM thus $G_{1,2} = 1$. The output schedule in Figure~\ref{fig:motivation}f has two tiles. In tile $\mathcal{T}_{0,1} = \{ 5,6, 6 \}$, $\{5,6\} \in I_{0,1}$ and $\{6\} \in J_{0,1}$.

\subsubsection{\textbf{Step 1}}
The first step of the tile fusion scheduler creates an intermediate fused schedule $\mathcal{F}$, composed of uniform coarse fused tiles to maximize the fused ratio while ensuring the load balance constraint. The scheduler first finds fused iterations from tiles of consecutive iterations to improve spatial locality and reduce the scheduler overhead. The scheduler also ensures iterations in different tiles of a wavefront are independent, no synchronization is needed.

Lines~\ref{lin:alg:sched:step1begin}--\ref{lin:alg:sched:step1end} in Algorithm~\ref{alg:scheduler} shows how the intermediate fused tiling $\mathcal{F}$ is created. The scheduler first computes the uniform tile size of $t$ using the given coarse tile size $ctSize$ in line~\ref{lin:alg:sched:step1begin}. As shown, the tile size is chosen to be $ctSize$ if the number of tiles, i.e., $\lceil|I|/ctSize\rceil$ is larger than or equal to $p$ otherwise, it defines $t=|I|/p$. 
This ensures the number of tiles in each wavefront is larger than $p$, i.e., the load balance constraint. Each fused tile $\mathcal{F}_{0,k}$ is created from $t$ consecutive iterations of $I$ as shown in line~\ref{lin:alg:sched:step1range} and some of $t$ consecutive iterations of $J$ as shown in line~\ref{lin:alg:sched:step1jloop}--\ref{lin:alg:sched:step1end}. An iteration of $J$ is added to tile $\mathcal{F}_{0,k}$ if all of its incoming edges are already in the tile as shown in line~\ref{lin:alg:sched:step1correctness}. Iterations that do not satisfy the criteria in line~\ref{lin:alg:sched:step1correctness} are added to tile $\mathcal{F}_{1,k}$ in the second wavefront as shown in line~\ref{lin:alg:sched:step1wf1}. The iterations in the second wavefront, $\mathcal{F}_1$ is evenly distributed into $t$ tiles using the $balance$ routine in line~\ref{lin:alg:sched:step1end} to ensure load balance in the second wavefront. 

\begin{figure}
    \centering
    \includegraphics[width=0.6\linewidth]{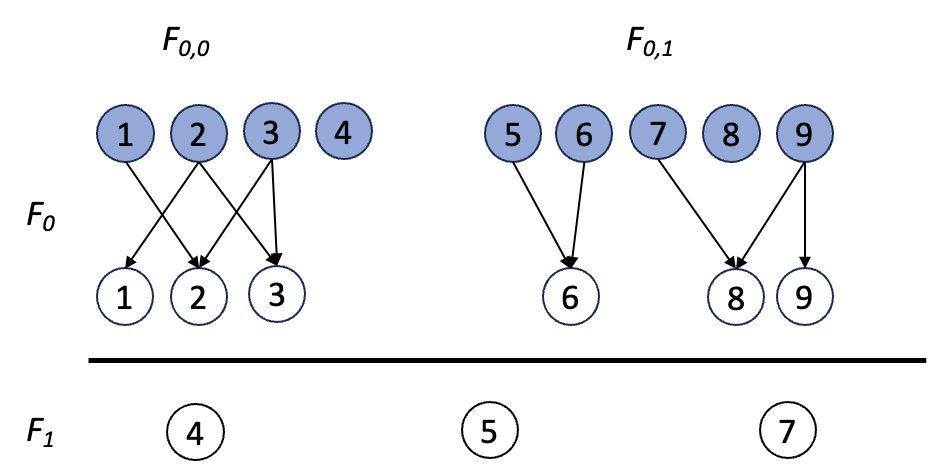}
    \caption{Tile fusion schedule after step 1}
    \label{fig:step1}
\end{figure}

\begin{figure}
    \centering
    \includegraphics[width=0.7\linewidth]{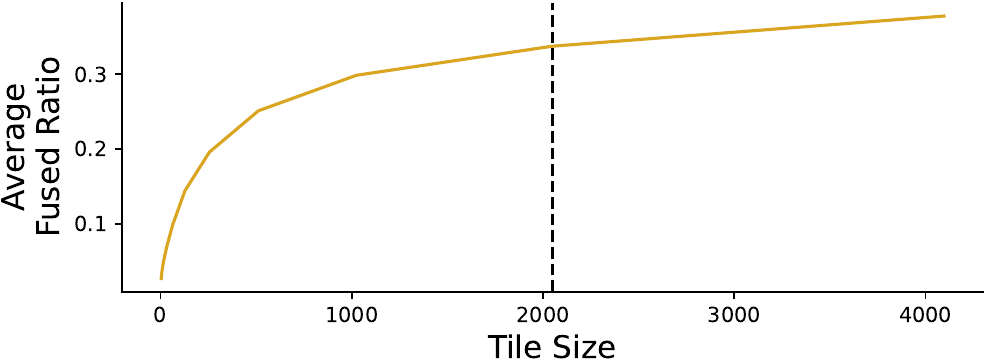}
    \caption{Variation of fused ratio versus tile size.}
    \label{fig:heuristic}
\end{figure}

The coarse tile size parameter, $ctSize$ used for specifying $t$ in line~\ref{lin:alg:sched:step1begin} in Algorithm~\ref{alg:scheduler}, is determined heuristically. To select the best value for $ctSize$, we compute how the fused ratio changes when tile size increases. Figure~\ref{fig:heuristic} shows tile size changes on the x-axis and the average of fused ratio changes for all matrices of the SuiteSparse repository on the y-axis. The value of $ctSize$ should be selected to maximize the tile fusion objective. Since after $ctSize=2048$ in Figure~\ref{fig:heuristic}, the rate of fused ratio improvement is slowed down, we use $ctSize=2048$. While going beyond this value can slightly increase the fused ratio, it reduces the number of tiles in a wavefront, potentially leading to load imbalance.

Figure~\ref{fig:step1} shows the output of step 1 for the example shown in Figure~\ref{fig:motivation}. For this example, we assume $ctSize=4$ and $p=3$ which makes a tile size $t=4$. Two coarse tile size is shown in Figure~\ref{fig:step1} $\mathcal{F}_{0,0} = {1,2,3,4}$ and then since iterations ${1, 2, 3} \in J$ depend on iterations ${1, 2, 3} \in I$ and they already exist in $\mathcal{F}_{0,0}$, then the three iterations are added.

\subsubsection{\textbf{Step 2}}
The second step of the tile fusion scheduler splits coarse tiles created in the first step, $\mathcal{F}$, to fit them into fast memory. As a result, tiles with different sizes are created to ensure the locality constraint of the tile fusion. The scheduler iterates over $\mathcal{F}$ and measures their data movement with a data movement cost model. Fused tiles whose data movement is larger than the size of fast memory are split to fit into the fast memory. 

The second step of the tile fusion scheduler is shown in Lines~\ref{lin:alg:sched:step2beg}--\ref{lin:alg:sched:step2end} in Algorithm~\ref{alg:scheduler}. 
The algorithm iterates over all tiles in the two wavefronts $\mathcal{F}_0$ and $\mathcal{F}_1$ and computes the data movement cost of each tile using a $cost$ function as shown in line~\ref{lin:alg:sched:step2cost}. If the data movement cost of a tile $\mathcal{F}_{i,j}$ is larger than the size of fast memory $cacheSize$, the scheduler splits the tile recursively to create a set of tiles, each fitting into the fast memory using the $split$ routine as shown in line~\ref{lin:alg:sched:step2split}. The resulting split tiles are added to the same wavefront of the final schedule $\mathcal{T}$.

\subsubsection*{Data movement cost}
The tile fusion scheduler relies on a cost model to approximate data movement for a coarse tile. As shown in line~\ref{lin:alg:sched:step2cost} in Algorithm~\ref{alg:scheduler}, the data movement cost model computes the cost of a tile $\mathcal{T}_{i,j}$ for a given $bCol$ and $cCol$. The cost model is computed as shown in Equation~\ref{eq:costmodel}:
    \begin{multline}
    \label{eq:costmodel}
    cost(\mathcal{T}_{i,j}, bCol, cCol) = \\ (nz(\mathcal{T}_{i,j}) +  uc(\mathcal{T}_{i,j}) + t + |J_{i,j}|) * cCol + idx 
    \end{multline}
where, $nz(\mathcal{T}_{i,j})$ is the number of unique nonzeros in the tile from $A$ and $B$. When $B$ is dense, all $n\times bCol$ is added. $uc(\mathcal{T}_{i,j})$ is the number of nonzeros with unique columns in the tile, $|J_{i,j}|$ is the number of fused iterations from the second operation and $idx$ is the indexing cost for when $A$ or $B$ are sparse.

The final fused schedule $\mathcal{T}$ in Figure~\ref{fig:motivation}f is computed from the coarse fused tile schedule in $\mathcal{F}$ in Figure~\ref{fig:step1}. The schedule has two coarse fused tiles. Each tile is labeled with its data communication cost. Assuming $bCol=cCol=1, t=4$, the cost of $\mathcal{F}_{0,0}$ is computed based on Equation~\ref{eq:costmodel}. 
Since the cost of $\mathcal{F}_{0,0}$ is less than $cacheSize=30$, the tile will be directly added to $\mathcal{T}$. However, $\mathcal{F}_{0,1}$ is larger than $cacheSize$ and is split in two smaller tiles that has a cost less than $cacheSize$.

\subsubsection*{\textbf{Computational Complexity}}
The first step of the algorithm, for every tile with size $t$, it checks $inEdges$ for only $t$ columns of $G$ in the same range. Since tiles are disjoint, $inEdges$ is called once per iteration, makes the first step of the algorithm $O(nnz)$. Accessing incoming edges for an iteration is possible in linear time.
The second step only iterates over fused iterations in $J$ and their dependent iterations from $I$ for splitting. Since the set of iterations is split by a factor of 2, and each split function can visit up to $nnz$ edges, therefore, its complexity is $O(nnz*log(ctSize)$. 
The second wavefront only operates on unfused iterations which take $O(|J|)$. Therefore the complexity of the second step is $O(|J| + nnz*log(ctSize)$.

\subsection{Fused Code}
The fused code is created by fusing the outermost loop of the two operations. The fused code is then replaced with a doubly nested loop that iterates over the tiles in parallel using the OpenMP scheduler.
The fused code uses the fused tiling schedule $\mathcal{T}$ and maps iterations of fused tiles to their associated code version to ensure correctness. Listing~\ref{lst:fused_gemm_spmm} and Listing~\ref{lst:fused_spmm_spmm_csr} show the fused code for GEMM-SpMM (Figure~\ref{fig:motivation}) and SpMM-SpMM (Listing~\ref{lst:spmm_spmm_csr}). As shown the fused code is composed of a version of the two operations. Lines~\ref{lin:gemmb}--\ref{lin:gemme} and \ref{lin:spmmb}--\ref{lin:spmme} in Listing~\ref{lst:fused_gemm_spmm} correspond to the innermost loops of GEMM and SpMM respectively. Lines~\ref{lin:spmm1b}--\ref{lin:spmm1e} and \ref{lin:spmm2b}--\ref{lin:spmm2e} in Listing~\ref{lst:fused_spmm_spmm_csr} correspond to the innermost loops of the first SpMM ($D_1=AC$) and the second SpMM ($D=AD_1$) respectively. Loop-bounds in line~\ref{lin:gemmb} in Listing~\ref{lst:fused_gemm_spmm} and line~\ref{lin:spmm1b} in Listing~\ref{lst:fused_spmm_spmm_csr} are determined by the schedule $\mathcal{T}$, playing as a runtime check to switch between the two versions. Therefore, the fused code follows the schedule order and ensures all dependence between fused iterations. The outermost loop of the two computations is fused and replaced with a pair of loops that go over the tile fusion schedule.

Fused code also turns data reuse between fused iterations into temporal locality while also preserving the locality of each operation. The code versions are next to each other in the code and when they execute right after each other, the data reuse between them turns into temporal locality. For the first tile in the schedule shown in Figure~\ref{fig:motivation}, the arrays corresponding to rows 1--4 stay in the cache when the fused code in Listing~\ref{lst:fused_gemm_spmm} switches to the SpMM loop in line~\ref{lin:spmmb}, thus improving temporal locality.  
Both fused codes in Listings~\ref{lst:fused_gemm_spmm} and \ref{lst:fused_spmm_spmm_csr} benefit from the fact that consecutive iterations are grouped in the scheduler. The consecutive choice will eliminate the need for conditional checking for every iteration of computations. It also improves spatial locality and temporal locality exist within iterations of each operation, e.g. spatial and temporal locality in GeMM will be in place in fused tiles.

\begin{lstlisting}[label={lst:fused_gemm_spmm},mathescape=true,style=cpp,
caption={Fused code of GeMM-SpMM, $D = A(BC)$. }]
for (w in T){ // for each wavefront
#pragma omp parallel for |\label{lin:ompPragma}|
 for(t in T[w]){ // for each tile
  for(i1 in t.first)|\label{lin:gemmb}|
   for(int i2=0; i2<cCol; ++i2)
    for(int i3=0; i3<bCol; ++i3)
     D1[i3][i2] += B[i1][i3] * C[i2][i3]; |\label{lin:gemme}|
  for(j1 in t.second) |\label{lin:spmmb}|
   for(int j2=A.p[j1]; j2<A.p[j1+1]; j2++)
    for(int j3=0; j3<cCol; j3++)
     D[j1][l2] += A.x[j2] * D1[A.i[j2]][j3]; |\label{lin:spmme}|
 }} 
\end{lstlisting}

\begin{lstlisting}[label={lst:spmm_spmm_csr},mathescape=true,style=cpp,
caption={ $D = A (A C) $ where $A$ is CSR.}, keywords={fuse} ]
fuse: for(int i1=0; i1<n; ++i1) // SpMM 1
  for(int i2=A.p[i1]; i2<A.p[i1+1]; ++j1) 
   for(int i3=0; i3<cCol; ++i3) 
    D1[i1][i3] += A.x[i2] * C[A.i[i2]][i3];
fuse: for(int j1=0; j1<n; ++j1) // SpMM 2
  for(int j2=A.p[j1]; j2<A.p[j1+1]; ++j2) 
   for(int j3=0; j3<cCol; ++j3) 
    D[j1][j3] += A.x[j2] * D1[A.i[j2]][j3];
\end{lstlisting}

\begin{lstlisting}[label={lst:fused_spmm_spmm_csr},mathescape=true,style=cpp,
caption={ Fused code for SpMM-SpMM, $D = A  (AC)$.} ]
for (w in T){
#pragma omp parallel for |\label{lin:ompPragmaSpMM}|
 for(t in T[w]){ |\label{lin:spmm1b}|
  for(i1 in t.first)
   for(int i2=A.p[i1]; i2<A.p[i1+1]; ++i2)
    for (int i3 = 0; i3<cCol; ++i3) |\label{lin:innerSpmm1}|
      D1[i1][i3] += A.x[i2] * C[A.i[i2]][i3];|\label{lin:spmm1e}|
  for(j1 in t.second) |\label{lin:spmm2b}|
   for(int j2 = A.p[j1]; j2<A.p[j1+1]; j2++)
    for(int j3 = 0; j3<cCol; j3++) |\label{lin:innerSpmm2}|
      D[j1][j3] += A.x[j2] * D1[A.i[j2]][j3];|\label{lin:spmm2e}| 
 }}
\end{lstlisting}

 The fused code enables thread-level parallelism by mapping fused tiles to parallel threads. Mapping tiles to threads is done using \texttt{omp} scheduler as shown in line~\ref{lin:ompPragma} in Listing~\ref{lst:fused_gemm_spmm} and line~\ref{lin:ompPragmaSpMM} in Listing~\ref{lst:fused_spmm_spmm_csr}. The fused code will keep all fine-grain parallelism opportunities such as vectorization that exist in the unfused code. For example, lines~\ref{lin:gemmb}--\ref{lin:gemme} in Listing~\ref{lst:fused_gemm_spmm} is mapped to a highly optimized GEMM BLAS to benefit from vector processors. Similarly, inner loop vectorization is performed in lines~\ref{lin:innerSpmm1} and \ref{lin:innerSpmm2} in Listing~\ref{lst:fused_spmm_spmm_csr}.

\section{Experimental Results}
This section discusses the performance of tile fusion with existing fused and unfused implementations for sparse matrices across two different shared memory processors. 
Overall tile fusion is faster than unfused and best-fused code with a geomean speedup of 1.98$\times$ and 3.52$\times$ and is scalable to 40 and 64 cores. 

\subsection{Setup}

\begin{table}
\small
    \centering
    \begin{tabular}{c||c|c}
        \hline
        Platform &  CascadeLake & EPYC \\
        \hline
         \# of sockets $\times$ cores &2 $\times$ 20 cores  & 2 $\times$ 32 cores \\
         \hline
         L1/L2/L3 Cache Sizes  & 32K/1M/28M  & 32K/512K/256M  \\
         \hline
         Compiler & ICC 2022  & GCC v.11 \\
         \hline
         BLAS & MKL BLAS 2022~\cite{wang2014intel} & BLIS~\cite{van2015blis} \\
    \end{tabular}
    \caption{Platform details}
    \label{tab:target}
\end{table}

\subsubsection{Environment}
All experiments are executed on multi-core processors shown in Table~\ref{tab:target} to show the performance of tile fusion cross-platform. The experiments are done in single-node except stated otherwise. Since both computations are used in GNN~\cite{zhou2020graph} and sparse iterative linear solvers with multiple right-hand side~\cite{aggarwal2021batched}, we test tile fusion for double-precision (DP) and single-precision (SP) data types, commonly used in machine learning and scientific computing respectively.  All experiments are tested for three different numbers of columns, 32, 64, and 128 in $B$ to enable testing different matrix sizes $B$ and $C$. Each reported time in the paper is the median of 7 runs. For each matrix, the theoretical FLOPs for the unfused code is computed and used for all implementations. Parameter $cacheSize$ in Algorithm~\ref{alg:scheduler} is set to the sum of L1+L2+(L3/cores) cache sizes in row 2 in Table~\ref{tab:target}. A close thread binding is selected and each thread is pinned to a physical core. 


\subsubsection{Matrix Dataset} We select 233 matrices from SuiteSparse~\cite{davis2011university} matrix repository to evaluate GeMM-SpMM and SpMM-SpMM computations. We select two groups of matrices to represent scientific computing and machine learning applications. \textit{I.} all 132 symmetric positive definite matrices larger than $10^5$ nonzero values. \textit{II.} all 111 square matrices related to graph applications larger than $10^5$ nonzeros with either integer or real types. We define a matrix as graph-related if a``graph'' keyword is included in its metadata.

\begin{figure*}[!h]
  \centering
    \includegraphics[width=\textwidth]{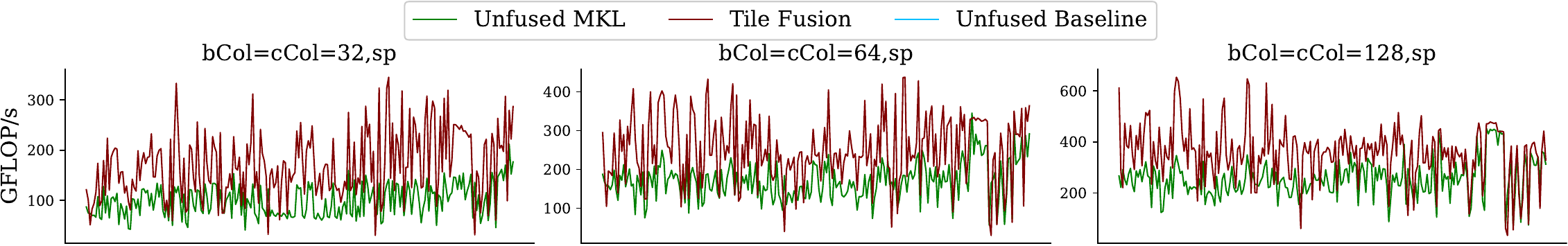} 
    \includegraphics[width=\textwidth]{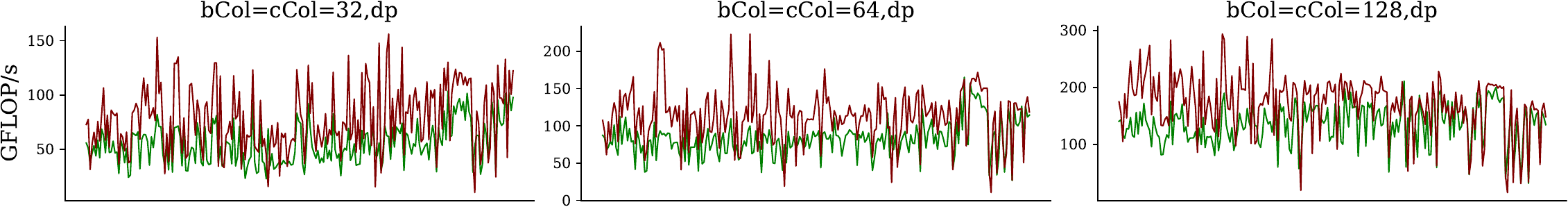}
    \includegraphics[width=\textwidth]{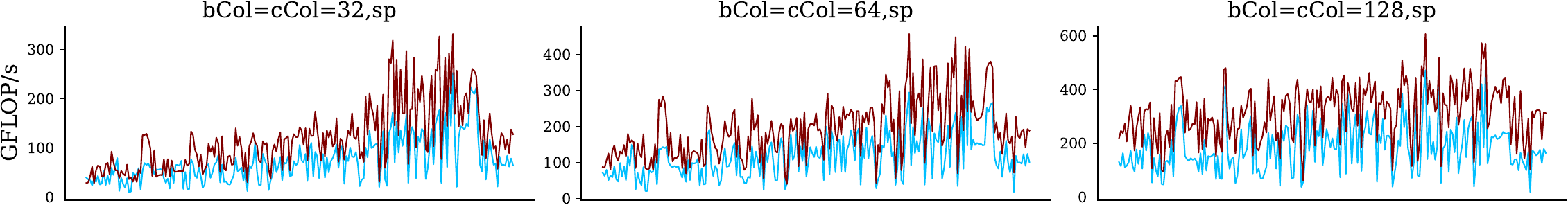} 
    \includegraphics[width=\textwidth]{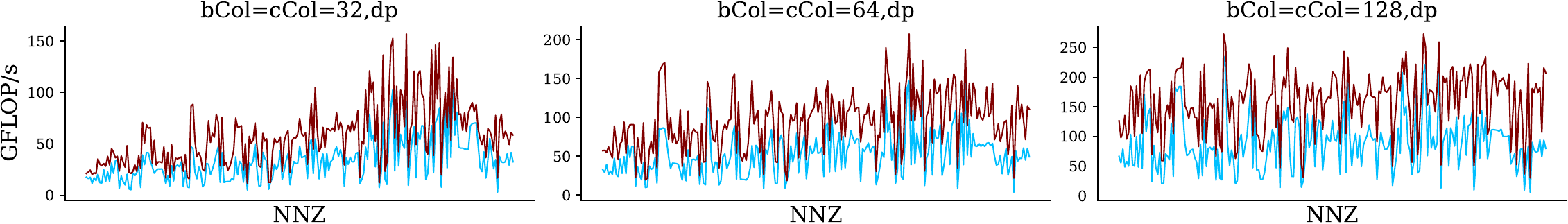} 
  \caption{GeMM-SpMM performance for all matrices on CascadeLAke (top) EPYC (bottom)}
  \label{fig:gemm_spmm_performance_unfused}
\end{figure*}

\subsubsection{Fused and Unfused Implementations}

To compare with unfused implementations, we use OneAPI MKL~\cite{wang2014intel} library v2022. Since fused versions of GEMM or SpMM are not supported in MKL, we called routines \texttt{cblas\_?gemm} and \texttt{mkl\_sparse\_?\_mm} for double/single precision of GeMM and SpMM, respectively. We set the number of threads in MKL with \texttt{mkl\_set\_num\_threads()} to the number of physical cores. We also develop an unfused parallel implementation for both GeMM-SpMM and SpMM-SpMM with the same set of optimizations to show the effect of tile fusion.

Tile fusion is compared with existing fused implementations and some in-house implementations of prior fusion techniques. For GeMM-SpMM, we use the generated \texttt{C++} code from TACO~\cite{kjolstad2017tensor} and SparseLNR~\cite{dias2022sparselnr} tensor compilers for the expression \texttt{D(i,l) = A(i,j) * B(j,k) * C(k,l)} where $A$ is sparse and other matrices are dense. We added both generated codes and reported their best timing as \texttt{Best of Tensor Compilers}. We also additionally vectorize the generated tensor compiler code by using MKL GeMV BLAS~\cite{wang2014intel} to ensure the effect of vectorization is incorporated. 

For SpMM-SpMM, tensor compilers do not support fusing SpMM-SpMM and, therefore excluded from the benchmark. Since the code for communication-avoiding~\cite{demmel2008avoiding} and sparse tiling~\cite{krieger2013loop} are not publicly available, per the authors' recommendation, we adopted the idea and applied them to SpMM-SpMM.  For communication-avoiding methods, we first equally partition iterations of the first SpMV, and then we add all dependent iterations to the same partitions. This implementation is called \textit{overlapped tiling}.  For sparse tiling, we partition iterations of the first SpMM equally. Then we add dependent iterations of the second SpMM to each partition to create balanced partitions. Finally, dependence is resolved with atomic and barriers. This implementation is named \textit{atomic tiling}.

We only report the fused code execution time for all experiments operating on matrices. 
The scheduler overhead is separately evaluated and illustrated. 

\begin{table}[!t]
\small
  \centering
  \begin{tabular}{|c |c | c c c c|} 
   \hline
    & &  & 32 & 64 & 128 \\ [0.5ex] 
   \hline
   \multirow{4}{5em}{CascadeLake} & \multirow{2}{7em}{Single Precision} & MKL & 1.64 & 1.41 & 1.36 \\
    & & UnFused & 1.36 & 1.24 & 1.14 \\
   \cline{2-6}
    & \multirow{2}{7em}{Double Precision} & MKL & 1.37 & 1.33 &  1.23 \\
    & & UnFused & 1.45 & 1.34 & 1.24 \\
   \hline
   \multirow{2}{2em}{EPYC} & Single Precision & UnFused & 1.67 & 1.73 &  1.84 \\
   \cline{2-6}
   & Double Precision & UnFused & 1.81 & 1.93 &  1.97  \\
   \hline
  \end{tabular}
  \caption{The summary of gmean of speedups for GeMM-SpMM.}
  \label{tab:speedups_gemm_spmm}
\end{table}
\begin{figure}[]
  \centering
    \includegraphics[width=\linewidth]{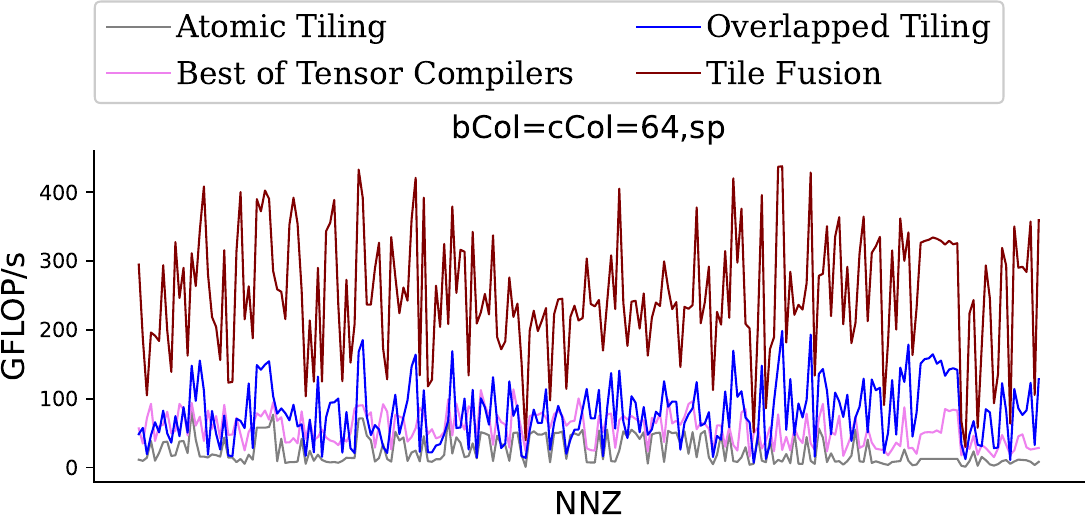} 
  \caption{GeMM-SpMM performance of fused implementations on graph matrices.}
  \label{fig:gemm_spmm_performance_fused}
\end{figure}

\subsection{GEMM-SpMM Evaluation}






\subsubsection{Performance in FLOPs}

Figure~\ref{fig:gemm_spmm_performance_unfused} shows the overall performance of GeMM-SpMM using tile fusion with unfused MKL for the two architectures and three bcols. As shown tile fusion is faster than MKL for 90\% of matrices across bclos. Table~\ref{tab:speedups_gemm_spmm} shows speedup details for GeMM-SpMM and for single and double precision for the target architectures shown in Table~\ref{tab:target}.

The performance of tile fusion increases as bCols increase due to increasing arithmetic intensity. The tile fusion performance increases from a mean of 152 GFLOP/s when bCol=32 to 328 GFLOP/s when bCol=128. While MKL implementation changes from 92 GLOP/s to 241 GFLOP/s when bCols changes from 32 to 128. As bCols increase, the arithmetic intensity of fused tiles increases and tile fusion can take advantage.

All implementations have a better performance for SPD matrices than graph matrices. The reason is that the fused ratio in SPD matrices is on average 2 times higher than graph matrices. 
%
The performance of Tile Fusion for single precision is 2$\times$ better than double precision. When operating on double, the data movement increases, making computation more memory-bound than single, thus reducing GFLOP/s. Also, since the EPYC processor has a larger L3 cache, the performance gap between tile fusion and unfused baseline for large matrices is higher than the CascadeLake processor.  

Tile fusion also supports fusing Equation~\ref{eq:chain} when the transpose of $C$ should be multiplied. Tile fusion provides a geometric mean of 1.49, 1.24, and 1.26 over unfused MKL on CascadeLake for bCol=cCol=32, 64, 128, respectively. 

Figure~\ref{fig:gemm_spmm_performance_fused} shows the performance of tile fusion compared to other fused implementations. Tile fusion is faster than tensor compilers, atomic tiling, and overlapped tiling with an average speedup of 9.4$\times$, 13.6$\times$, and 3.5$\times$, respectively. Tensor compilers perform redundant computations and also do not use memory hierarchy due to vector operations.

\subsubsection{Ablation Study}
This section analyzes the effect of tile fusion on locality and load balance and the effect of the two steps of the tile fusion scheduler on the performance. We selected all 111 graph matrices, a subset of the matrix dataset for profiling and analysis. All analysis is also done on the CascadeLake target architecture.

\begin{figure}
  \centering
    \includegraphics[width=0.8\linewidth]{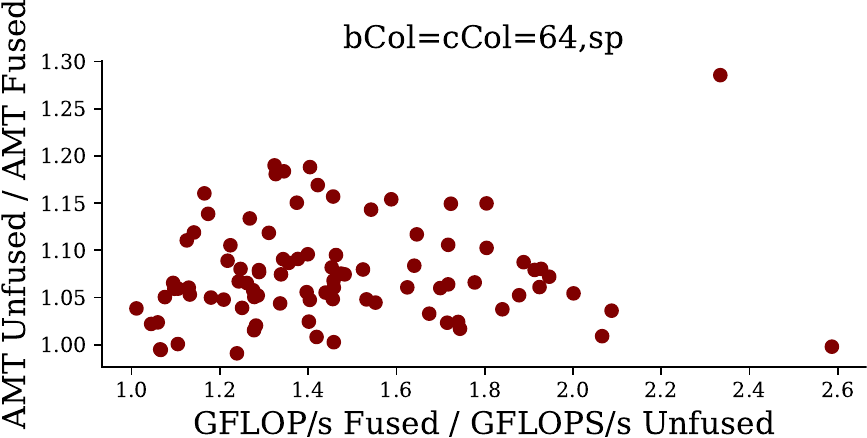} 
  \caption{The effect of average memory access time on performance of tile fusion for GeMM-SpMM}
  \label{fig:gemm_spmm_mem_cycle}
\end{figure}
We measure an average memory cycle to analyze the effect of tile fusion on improving locality in GeMM-SpMM. We measure average memory access time (AMT) as \textit{AMT = hit time + miss ratio * miss penalty} for all three levels of caches in the target architecture. We use PAPI~\cite{terpstra2010collecting} performance counters, \texttt{PAPI\_L1\_TCM}, \texttt{PAPI\_L2\_TCM}, \texttt{PAPI\_L3\_TCM} to measure L1 accesses, L2 accesses, L3 accesses, and main memory accesses, respectively to compute hit and miss ratio for each level. Average memory access times for the selected subset of matrices are shown in Figure~\ref{fig:gemm_spmm_mem_cycle}.  As shown, tile fusion improves AMT for 92\% of graph matrices between 1.1-1.3$\times$ compared to the unfused implementation which is the main cause for improving the performance.

\begin{figure}
  \centering
    \includegraphics[width=0.8\linewidth]{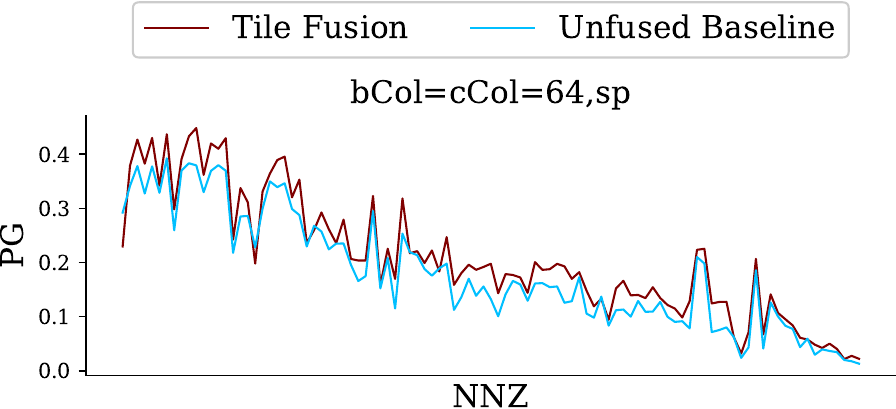} 
  \caption{Potential Gain of GeMM-SpMM (lower is better) }
  \label{fig:gemm_spmm_pg}
\end{figure}

We measure the potential gain of both fused and unfused code to show the effect of tile fusion on the load balance of GeMM-SpMM. Potential gain (PG) is defined as the maximum time that can be saved if all threads are balanced. We measure the average difference between the maximum time of threads and other threads' time. We use PAPI counter \texttt{PAPI\_TOT\_CYC} to measure the number of cycles for each thread. Figure~\ref{fig:gemm_spmm_pg} shows the PG compared to unfused. As shown, the tile fusion load balance is close to unfused. The unfused code has a larger number of fine-grain tasks, enabling it to be more balanced.   


\begin{figure}
  \centering
    \includegraphics[width=\linewidth]{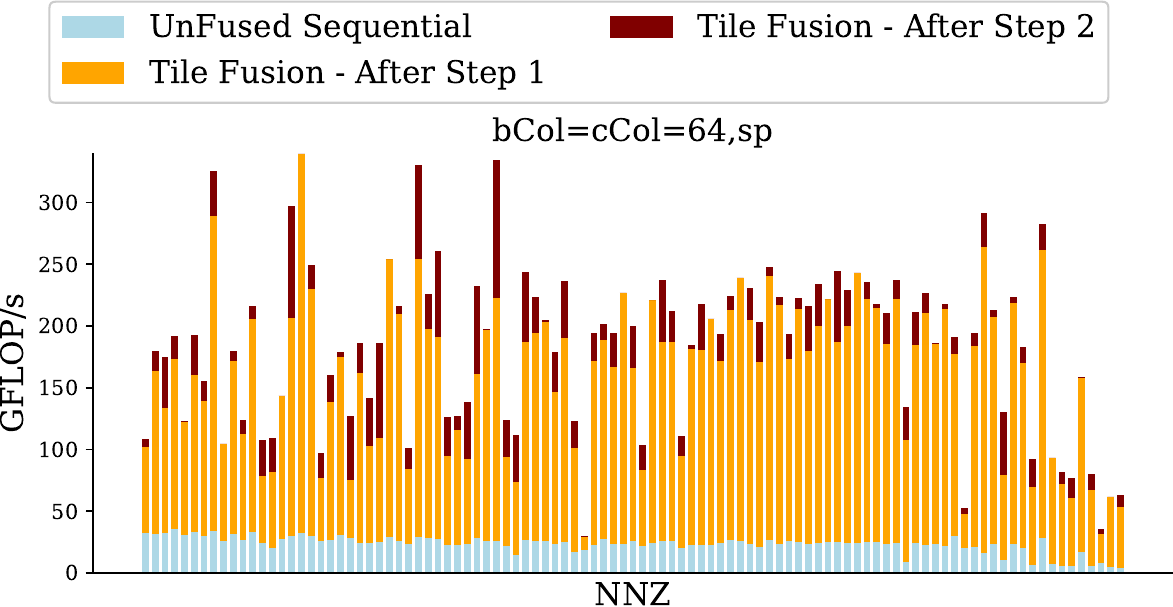} 
  \caption{Effect of different steps of Tile Fusion in GeMM-SpMM for graph matrices }
  \label{fig:sgemm_spmm_stacked}
\end{figure}


Figure~\ref{fig:sgemm_spmm_stacked} shows the performance breakdown of the two steps of the tile fusion inspector. As shown, the first step of tile fusion improves the performance of sequential baseline code with a gmean speedup of 6.7$\times$. The second step of tile fusion contributes to the performance of 90\% of matrices shown in Figure.  This first step contributes more because it adds threading and improves locality. The second step further balances the loads and improves the parallel workloads of step 1. The second step selects tile sizes based on the cost model provided in Equation~\ref{eq:costmodel}. For the selected graph matrices, the tile sizes selected by the second step vary between 64-2048.  

\begin{figure}[!t]
  \centering
    \includegraphics[width=\linewidth]{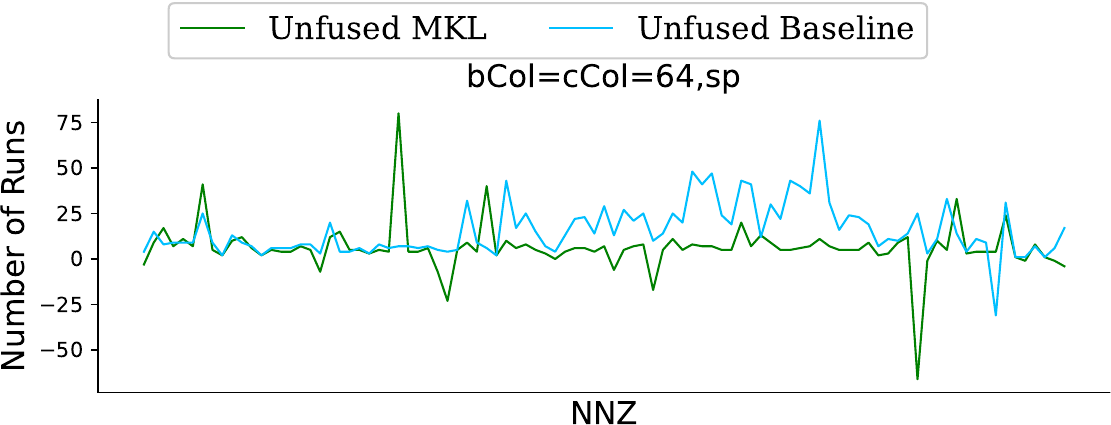} 
  \caption{Number of runs needed to amortize the scheduling cost for GeMM-SpMM (lower and positive values are better)}
  \label{fig:gemm_spmm_analysis}
\end{figure}

\begin{figure*}[!ht]
  \centering
  \begin{tabular}{c}
    \includegraphics[width=\textwidth]{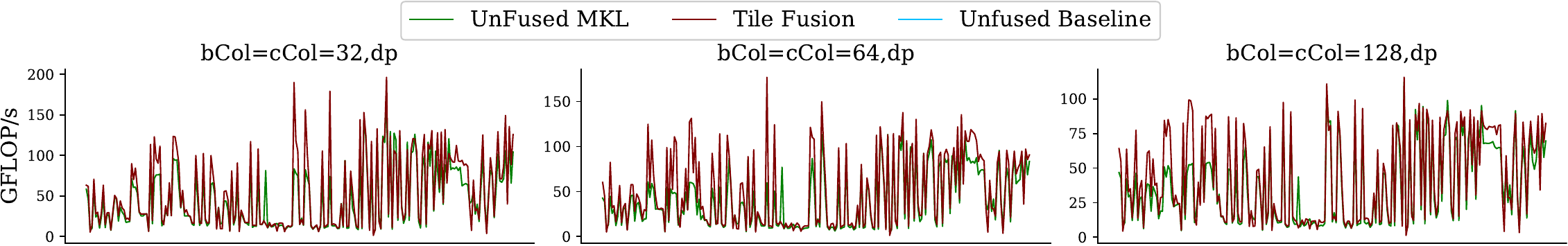} \\
    \includegraphics[width=\textwidth]{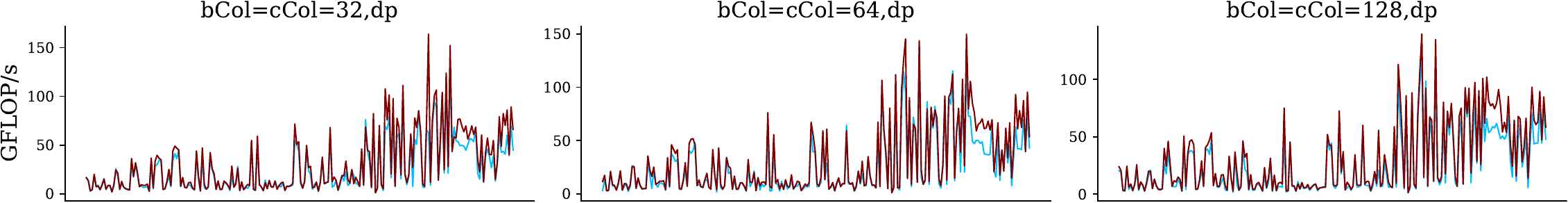} \\
  \end{tabular}
  \caption{SpMM-SpMM performance on CascadeLake (top) and EPYC (bottom)}
  \label{fig:spmm_performance_all_graphs}
\end{figure*}
\begin{figure}
  \centering
    \includegraphics[width=\linewidth]{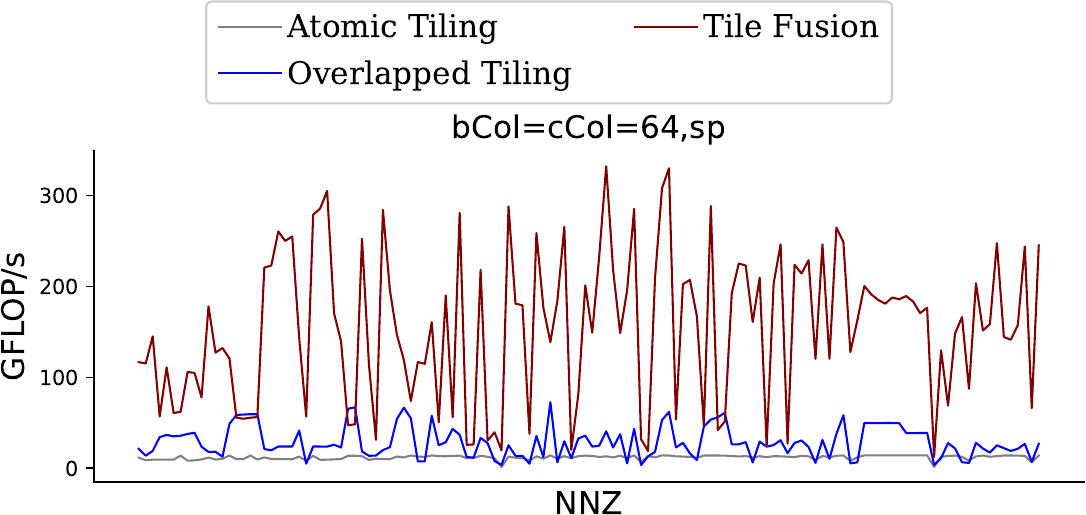} 
  \caption{SpMM-SpMM performance of fused implementations}
  \label{fig:spmm_prior_graphs}
\end{figure}

\begin{table}[!t]
\small
  \centering
  \begin{tabular}{|c |c | c c c c|} 
   \hline
    &  &  & 32 & 64 & 128 \\ [0.5ex] 
   \hline
   \multirow{4}{5em}{CascadeLake} & \multirow{2}{7em}{Single Precision} & MKL & 1.2 & 1.02 & 1.11 \\
   & & UnFused & 1.17 & 1.15 &  1.14 \\
   \cline{2-6}
   & \multirow{2}{7em}{Double Precision} & MKL & 1.09 & 1.16 &  1.11 \\
   & & UnFused & 1.14 & 1.15 &  1.13 \\
   \hline
   \multirow{2}{2em}{EPYC} & Single Precision & UnFused & 1.14 & 1.17 &  1.19 \\
   \cline{2-6}
   & Double Precision & UnFused & 1.14 & 1.20 & 1.22 \\
   \hline
  \end{tabular}
  \caption{G-mean speedups for SpMM-SpMM}
  \label{tab:speedupspmm}
\end{table}



\subsubsection{Scheduler Overhead analysis}
The tile fusion performs scheduling once per sparsity pattern and can be reused as long as sparsity remain static. Figure~\ref{fig:gemm_spmm_analysis} shows the number of iterations that fused code should run to amortize the scheduler overhead with respect to the fastest baselines. The number of fused code runs is computed as $\frac{scheduler \, time}{baseline-fusedCode \, time}$. As shown, tile fusion needs less than 100 iterations to amortize the cost of the scheduler. In many applications such as GNN training, GeMM-SpMM is called over hundreds or thousands of times.

\subsection{SpMM-SpMM Evaluation}






The performance of tile fusion is compared with unfused implementations for SpMM-SpMM as shown in Figure~\ref{fig:spmm_performance_all_graphs}. Tile fusion is faster than unfused baseline and MKL implementations in 100\% and 70\% of all matrices in any bCol that we experimented on and for SP/DP. The detailed speedup for both CascadeLake and EPYC and SP and DP are illustrated in Table~\ref{tab:speedupspmm}.
The performance of SpMM-SpMM is overall lower than GeMM-SpMM for the same set of matrices due to the memory-bound nature of SpMM.


Tile fusion provides a gmean speedup of 9.3$\times$, 13.2$\times$, and 13.7$\times$ over atomic tiling for bCol = 32, 64, and 128 respectively. A similar trend exists for overlapped tiling where tile fusion provides a gmean speedup of 5, 6.5, and 7.2 for bcols=32, 64, and 128. The main reason is the amount of redundant computation that increases for overlapped tiles. For example, matrix G2\_circuit and inline\_1 have redundant iterations of 126487 and 2844351 respectively while they only have 150102 and 503712 rows. 

\section{Related Work}

\subsubsection*{Tiling and Fusion for Sparse Codes}
Loop tiling and fusion are common techniques to improve locality. A large body of work applies these techniques to SpMM~\cite{natesh2023rosko,hong2019adaptive,kurt2020efficient,ahmad2024exploring,cheshmi2022vectorizing,wilkinson2023register} or GeMM~\cite{wang2014intel,kung2021cake} routines. Tile fusion preserve locality between iterations of GeMM and SpMM thus enables benefiting from existing developed efficient GeMM and SpMM.  Using data communication cost to define tile size is used in matrix signature~\cite{kurt2020efficient} for SpMM CSR. Tile fusion however works on two operations that needs to take into account common elements between the two operations.     
Fusing based on compile-time information is commonly used in dense matrices. Indirect memory accesses reduce fusion opportunities however, there are still opportunities.
Tensor expression compilers~\cite{COMET:LCPC-20,dias2022sparselnr,kjolstad2017tensor} generates code for a tensor expression. Sparse tensor compilers~\cite{COMET:LCPC-20,dias2022sparselnr,kjolstad2017tensor} support fusing Equation~\ref{eq:chain}. However, the generated fused code from SparseLNR~\cite{dias2022sparselnr} and TACO~\cite{kjolstad2017tensor} make matrix operations in Equation~\ref{eq:chain} to matrix-vector operations which do not efficiently use the fast memory. 

Modeling parallel loops such consecutive matrix multiplication as a graph~\cite{krieger2013loop,strout2004sparse,demmel2008avoiding} or hypergraph~\cite{pawlowski2020combinatorial} are commonly used for improve cache reuse. These methods for shared memory processors either rely on synchronization~\cite{krieger2013loop} or overlapped computation~\cite{demmel2008avoiding}. The fused ratio and its cost model can enhance locality and reduce overlapped computation in these methods. Sympiler~\cite{cheshmi2022transforming,cheshmi2017sympiler} uses DAG schedulers~\cite{cheshmi2018parsy,zarebavani2022hdagg} to build an initial schedule of iteration then fuse the schedule with another loop using sparse fusion~\cite{cheshmi2023runtime,cheshmi2022optimizing}. Sparse fusion scheduler is driven by loop-carried dependency commonly occurs in scientific solvers~\cite{cheshmi2020nasoq} to ensure locality which does not exist in matrix multiplication operations. 

\section{Conclusion}
This paper presents tile fusion to enable fusing GeMM-SpMM and SpMM-SpMM for $D = A \times B \times C$ where $A$ is sparse and $B$, $C$, and $D$ are dense. Tile fusion has a scheduler that builds a schedule of fused tiles based on the sparsity pattern of the $A$ and the size of dense matrices. The created schedule does not use redundant computation and its synchronizations are always 2. Tile fusion outperforms existing unfused and fused implementations.

\bibliography{sample-base}


\end{document}